\documentclass[12pt,preprint]{aastex}

\newcommand{\bdv}[1]{\mbox{\boldmath$#1$}}

\def\au{{\rm au}}

\def\kms{{\rm km}\,{\rm s}^{-1}}
\def\masyr{{\rm mas}\,{\rm yr}^{-1}}
\def\kpc{{\rm kpc}}
\def\mas{{\rm mas}}

\def\muas{\mu{\rm as}}

\def\pc{{\rm pc}}

\def\rel{{\rm rel}}

\def\e{{\rm E}}

\def\bmu{{\bdv\mu}}

\def\btheta{{\bdv\theta}}

\begin{document}
\title{KMT-2017-BLG-2820 and the Nature of the Free-Floating Planet Population}

\author{\textsc{Yoon-Hyun Ryu$^{1}$, Przemek Mr{\'o}z$^{2}$, Andrew
Gould$^{3,4}$, Kyu-Ha Hwang$^{1}$, Hyoun-Woo Kim$^{1,5}$, Jennifer
C. Yee$^{6}$ \and  Michael D.Albrow$^{7}$, Sun-Ju Chung$^{1,8}$,
Youn Kil Jung$^{1}$, In-Gu Shin$^{1}$, Yossi Shvartzvald$^{9}$,
Weicheng Zang$^{10}$, Sang-Mok Cha$^{1,11}$, Dong-Jin Kim$^{1}$,
Seung-Lee Kim$^{1,8}$, Chung-Uk Lee$^{1,8}$, Dong-Joo Lee$^{1}$,
Yongseok Lee$^{1,11}$, Byeong-Gon Park$^{1,8}$, Cheongho Han$^{12}$, Richard W. Pogge$^{4}$ \\
(KMTNet Collaboration)\\
Andrzej Udalski$^{13}$, Radek Poleski$^{13}$, Jan Skowron$^{13}$,
Micha{\l} K. Szyma{\'n}ski$^{13}$, Igor Soszy{\'n}ski$^{13}$,
Pawe{\l} Pietrukowicz$^{13}$, Szymon Koz{\l}owski$^{13}$, Krzysztof
Ulaczyk$^{14}$, Krzysztof A. Rybicki$^{13}$, Patryk Iwanek$^{13}$\\
(OGLE Collaboration)\\} }

\affil{$^{1}$Korea Astronomy and Space Science Institute, Daejon
34055, Republic of Korea}

\affil{$^{2}$Division of Physics, Mathematics, and Astronomy,
California Institute of Technology, Pasadena, CA 91125, USA }

\affil{$^{3}$Max-Planck-Institute for Astronomy, K\"{o}nigstuhl 17,
D-69117 Heidelberg, Germany}

\affil{$^{4}$Department of Astronomy, Ohio State University, 140 W.
18th Ave., Columbus, OH 43210, USA}

\affil{$^{5}$Department of Astronomy and Space Science, Chungbuk
National University, Cheongju 28644, Republic of Korea}

\affil{$^{6}$ Center for Astrophysics $|$ Harvard \& Smithsonian, 60
Garden St., Cambridge, MA 02138, USA}

\affil{$^{7}$University of Canterbury, Department of Physics and
Astronomy, Private Bag 4800, Christchurch 8020, New Zealand}

\affil{$^{8}$Korea University of Science and Technology, Korea,
(UST), 217 Gajeong-ro, Yuseong-gu, Daejeon, 34113, Republic of
Korea}

\affil{$^{9}$IPAC, Mail Code 100-22, Caltech, 1200 E. California
Blvd., Pasadena, CA 91125, USA}

\affil{$^{10}$Physics Department and Tsinghua Centre for
Astrophysics, Tsinghua University, Beijing 100084, China}

\affil{$^{11}$School of Space Research, Kyung Hee University,
Yongin, Kyeonggi 17104, Republic of Korea}

\affil{$^{12}$Department of Physics, Chungbuk National University,
Cheongju 28644, Republic of Korea}

\affil{$^{13}$Astronomical Observatory, University of Warsaw,
Al.~Ujazdowskie~4, 00-478~Warszawa, Poland}

\affil{$^{14}$Department of Physics, University of Warwick, Gibbet
Hill Rd., Coventry, CV4~7AL,~UK}

\begin{abstract}

We report a new free-floating planet (FFP) candidate, KMT-2017-BLG-2820,
with Einstein radius $\theta_\e\simeq 6\,\muas$, lens-source
relative proper motion $\mu_\rel \simeq 8\,\masyr$, and Einstein
timescale $t_\e=6.5\,$hr.  It is the third FFP candidate
found in an ongoing study of giant-source finite-source point-lens (FSPL)
events in the KMTNet data base, and the sixth FSPL FFP candidate
overall.  We find no significant evidence for a host.
Based on their timescale distributions and detection rates,
we argue that five of  these six FSPL FFP candidates are
drawn from the same population as the six point-source point-lens (PSPL)
FFP candidates found by \citet{mroz17} in the OGLE-IV data base.
The $\theta_\e$ distribution of the FSPL FFPs implies that they are
either sub-Jovian
planets in the bulge or super-Earths in the disk.  However, the apparent
``Einstein Desert'' ($10\la\theta_\e/\muas\la 30$) would argue for the latter.
Whether each of the 12 (6 FSPL and 6 PSPL) FFP candidates  is truly an FFP,
or simply a very wide-separation
planet, can be determined at first adaptive optics (AO) light on 30m
telescopes, and earlier for some.
If the latter, a second epoch of AO observations could measure the
projected planet-host separation with a precision ${\cal O}(10\,\au)$.
At the present time, the balance of evidence favors the unbound-planet
hypothesis.

\end{abstract}


{\section{{Introduction}
\label{sec:intro}}

Formally, free-floating planets (FFPs) are planetary mass ($M<13\,M_J$)
objects that are not bound to any star.  However, from a theoretical
viewpoint, one would like to distinguish between objects that
formed in situ, as stars do via gravitational collapse, and those
that formed in protoplanetary disks, like planets, and were subsequently
ejected.  This can only be done statistically, and only by a method
that is sensitive to a broad range of masses that extends well below
 $M\la M_J$, i.e., objects that are non-luminous given current technology.
That is, gravitational microlensing is the only current technique
by which such studies can be carried out.

Because it is increasingly difficult to form low-mass objects by
gravitational collapse, and also increasingly difficult to form
high-mass objects in protoplanetary disks (and even more difficult
to then eject them),  one expects a ``gap'' (more accurately, a strong
minimum) between these two regimes, which is analogous to the
so-called ``brown dwarf desert''.  The appearance of such a ``gap''
would allow one to individually identify the objects that likely formed
within the protoplanetary disk, thus enabling further study.

In fact, the short-timescale, single-lens/single-source (1L1S) events
that are the expected signature of FFPs can also be generated by
planets in wide orbits.   For example, if a doppelganger of
our own solar system were oriented face-on and lay half way to the Galactic
center, then its ``Neptune'' would lie about 7.5 Einstein radii
from its ``Sun''.
Hence, for most trajectories of the lens system relative to a background
source, a microlensing event due to the ``Neptune''
would be indistinguishable from 1L1S,
even though the planet is bound.  Therefore, to distinguish between
wide-orbit planets (which are themselves quite interesting) and FFPs,
one must wait for the source and lens to be sufficiently displaced
that they can be separately imaged.  Because a putative host might be
very faint, while sources range from upper main-sequence stars to
giants, this means waiting until the separation is adequate to resolve
at severe to extreme contrast ratios.  We adopt a range of
1.2 -- 1.5 FWHM from main-sequence to giant sources.
With diffraction limited imaging, this requires waiting
\begin{equation}
\Delta t = 13\,{\rm yr}\times (1,1.25)
\biggl({\lambda\over 2.2\,\mu{\rm m}}\biggr)
\biggl({D\over 10\,{\rm m}}\biggr)^{-1}
\biggl({\mu_\rel\over 5\,{\rm mas\,yr^{-1}}}\biggr)^{-1} ,
\label{eqn:delay}
\end{equation}
where $\lambda$ is the wavelength of the observations,
$D$ is the diameter of the mirror, $\mu_\rel$ is the lens-source relative
proper motion, and 1.25 = 1.5/1.2.
For most FFP candidates discovered to date, and also for those that will be
discovered in data from the next few years, this means waiting
until the end of this decade, when the separations become accessible
to current telescopes and when adaptive optics (AO) imaging on
30m class telescopes becomes available (thereby reducing the pre-factor
in Equation~(\ref{eqn:delay}) by a factor $\sim 3$).

Thus, the study of FFPs is intrinsically a long-term project.

The first approach to the study of FFPs was based on the
distribution of Einstein timescales,
\begin{equation}
t_\e = {\theta_\e\over\mu_\rel},
\qquad
\theta_\e = \sqrt{\kappa M\pi_\rel},
\qquad
\kappa\equiv {4G\over c^2\,{\rm au}}\simeq 8.14\,{{\rm mas}\over M_\odot},
\label{eqn:tedef}
\end{equation}
of 1L1S events.  Here, $\theta_\e$ is the Einstein radius and $\pi_\rel$
is the lens-source relative parallax.
\citet{sumi11} found a strong excess of $t_\e\sim 1\,$day events,
which they interpreted as due to a population of roughly Jupiter-mass
planets, with about twice the frequency of stars.  However, using
an independent and superior data set, \citet{mroz17} ruled out such an excess.
Nevertheless, \citet{mroz17} found an excess of few-hour events, which they
suggested is consistent with a population of Earth- or super-Earth-mass
objects.

Of particular note in the present context, the \citet{mroz17}
excess was separated
from the main distribution by a clear gap\footnote{The gap appears
{\it despite} the fact that the timescale errors are fairly large, with
typical $1\,\sigma$ confidence intervals spanning a factor of two in $t_\e$},
whereas the \citet{sumi11} excess
was not.  This difference is not due to the different qualities of
the two data sets, but is simply the result of the respective
ratios of the mean timescale of the putative planets to that of the
bulge lenses from the bottom
of the stellar and brown dwarf population (say $0.01\,M_\odot$), i.e.,
$\langle t_\e \rangle \sim \sqrt{\kappa (0.01\,M_\odot)(16\,\muas)}/(5\,\masyr)
= 2.6\,$days.
For the putative \citet{sumi11} planets, this ratio was about 0.4.  Hence,
the short-timescale tail of the brown-dwarf (BD) distribution
strongly overlaps the peak of the FFP distribution due to the
wide range of values of $\pi_\rel$ and $\mu_\rel$ entering
Equation~(\ref{eqn:tedef}).
However, the corresponding ratio for the \citet{mroz17} excess is
only about 0.08, implying that the BD-tail is negligibly small.

Given that the Jupiter FFP population
is at least 8 times smaller than suggested by \citet{sumi11}, it can hardly be
studied at all using the global $t_\e$ distribution.  That is,
the Jupiter FFP events result in a small excess of $t_\e\sim 1\,$day
events, relative to the higher-mass ``background'', which is difficult to
detect even statistically.  Moreover, such a small excess could
in principle be due to imperfect modeling of the BD population,
or even to the long-timescale tail of lower-mass FFPs. By contrast,
there is essentially no ``background''  of unrelated microlensing
events for the few-hour
events found by \citet{mroz17}: the only real issues are whether these brief
``bumps'' in the light curve are really due to microlensing, and if so,
whether these ``isolated'' low-mass objects are due to FFPs rather
than wide-separation bound planets.  Regarding the first question,
\citet{mroz17} argued that these events are likely due to microlensing.

Regarding the second question, as mentioned above, one can in principle
wait for the lens and source to separate and then conduct AO imaging.
However, because these are point-source (PSPL) rather than
finite-source (FSPL) single lens events,
there is no measurement of $\mu_\rel$, and therefore
Equation~(\ref{eqn:delay}) does not give a definite estimate of
how long one must
wait.  If one sets a ``reasonably conservative'' lower limit\footnote{
For the adequate approximation that bulge stars have an isotropic Gaussian
proper motion distribution with standard deviation $\sigma=2.9\,\masyr$,
events with $\mu_\rel < \mu_{\rel,\rm lim}\rightarrow 1.5\,\masyr$
constitute a fraction
$f=(2/\sqrt{\pi})\int_0^{(\mu_{\rel,\rm lim/2\sigma})^2} dx\,\sqrt{x}e^{-x}
\simeq (4/3\sqrt{\pi})(\mu_{\rel,\rm lim/2\sigma})^3\rightarrow 1.3\%$
of all bulge-bulge microlensing events.  That is, in this regime,
$f\propto \mu_{\rel,\rm lim}^3$
}
$\mu_\rel > 1.5\,\masyr$, then with the fiducial parameters, one
should wait 43 years.  However, this still implies that most of the
\citet{mroz17}
FFP candidates can be vetted at, or soon after, first AO light on 30m
telescopes.

If some or all of the FFP candidates prove to be wide-orbit planets,
then they can be subjected to further study.  For each case, the
planet-host separation can be measured with a precision of about
$10\,\au$ as we discuss in Section~\ref{sec:discuss}.  Then, they can
be classified into either true analogs of Uranus and Neptune, which appear
to have been ``ejected'' from their Jupiter/Saturn-like orbits but remaining
in the same region of the Solar System, or those that have been
ejected into Kuiper-like or even Oort-like orbits \citep{gould16}.

A second approach to the investigation of FFPs was pioneered by
\citet{ob161540}, who searched for OGLE light curves that were consistent with
being due to short 1L1S microlensing events, but often with incomplete
coverage.  Then they checked whether the event could be characterized
using other survey data.  Although not specifically intended,
this approach tends to select FSPL events with source crossing
times $t_* \equiv \rho t_\e$ of about half a day or so.  Here,
$\rho=\theta_*/\theta_\e$ is the ratio of the source-star angular
radius to the Einstein radius.  Substantially
longer events would be adequately covered by OGLE itself and would
therefore not require this hybrid approach.  Substantially
shorter events would either fall mostly inside or mostly outside
a single night of OGLE data and therefore either would not require
this approach or would not be detected by it at all.  PSPL events
of similar effective duration have sufficient longer-term structure
to characterize them from several nights of data.  To date, this
approach has yielded two FSPL FFP candidates: OGLE-2016-BLG-1540
and OGLE-2012-BLG-1323 \citep{ob161540,ob121323}, with $t_* = 0.53\,$days and
$t_* = 0.78\,$days respectively.  In both cases, OGLE data
covered regions near the peak of the box-like light curve,
and these data had to be supplemented by data from other time zones
to be properly interpreted.  These were from Australia and South Africa
in the first case, and from Israel and New Zealand in the second.

OGLE discovered one other FSPL FFP directly from its Early Warning System
(EWS, \citealt{ews1,ews2}), OGLE-2019-BLG-0551 \citep{ob190551}.
Because its self-crossing time was
much longer, $t_* = 1.7\,$days, this event did not require any other
data for full characterization, although KMTNet \citep{kmtnet} data were
included
in the fit.  And a special OGLE search yielded the very short
FFP event, OGLE-2016-BLG-1928 \citep{ob161928},
which had not been alerted by OGLE EWS.
With $t_*=0.10\,$days, the event was almost completely contained
within only one night of OGLE data, although the
post-event (completely flat) KMTNet data
from South Africa were helpful in ruling out 2L1S alternative solutions.  Thus,
these two FFP candidates tend to confirm that the hybrid approach
of \citet{ob161540} tends to select FSPL events with $t_* \sim 0.5\,$days.
The source stars for such events are essentially all giants.

In the course of their 2019 annual review of microlensing events
found by their EventFinder system \citep{eventfinder}, KMTNet identified
KMT-2019-BLG-2073 as a likely FFP candidate.  \citet{kb192073} then
showed that, like the previous three\footnote{OGLE-2016-BLG-1928 had
not yet been discovered.} FSPL FFP events (defined as
having $\theta_\e \la 10\,\muas$), KMT-2019-BLG-2073 had a giant-star
source and $\rho\ga 1$.  In order of discovery, these four events
had $\theta_* = (12,15,20,5.4)\,\muas$ and
$\rho=(5.0,1.6,4.5,1.1)$.
This led \citet{kb192073} to suggest a systematic search for such
giant-source FSPL FFP events.  Moreover, they immediately recognized
that if such a search were extended to {\it all} giant-source FSPL
events, it would have substantially greater scientific value.  They
developed an automated algorithm for searching the KMT EventFinder
list for FSPL candidates, as well as procedures to vet these.  They
carried out an additional special search for giant-source events
based on a variant of EventFinder that would be more forgiving of
FSPL light curve distortions and more inclusive of giant sources
than the standard EventFinder.  In addition, more representations of
the light curve were shown to the operator than in the general search
to enable easier recognition of non-standard light curves.
\citet{kb192073} showed concretely that these
searches and procedures were tractable by applying them to the
2019 KMT data.  And they demonstrated that the results were statistically
well behaved.

\citet{kb192073} found a total of 13 FSPL events, of which two
were ``planetary'' ($\theta_\e < 10\,\muas$).  There was a ``gap''
of $\Delta\log\theta_\e = 0.82$ between the two FFPs and the next
smallest $\theta_\e$, while all of the 11 other increments in the cumulative
distribution function had $\Delta\log\theta_\e <0.28$.

While cautioning
that no statistical conclusions could be drawn about FFP frequency from the 2019
sample (due to publication bias), \citet{kb192073} argued that the
gap was likely real and, in any case, could be tested by carrying
out similar searches on other seasons of KMT data.  This illustrates
one of the powerful advantages of FSPL studies of FFPs: the corresponding
gap in the underlying $t_\e$ distribution of 1L1S events would be substantially
weaker.  There are two reasons for the difference.  First, from
Equation~(\ref{eqn:tedef}), the $\theta_\e$ distribution is less
``smeared out''  relative to the mass distribution because there
is only one degenerate variable $(\pi_\rel)$, rather than two
$(\pi_\rel,\mu_\rel)$ for the $t_\e$ distribution.  Second, the cross
section for 1L1S microlensing events is $\theta_\e\propto \sqrt{M}$,
whereas the cross section for FSPL events is $\theta_*$, which is
independent of lens mass.  Hence, the population of BD events that can
``scatter down'' to the planetary regime is suppressed for the FSPL
($\theta_\e$) distribution relative to the 1L1S ($t_\e$) distribution.

A second advantage is that the independent measurement of $\mu_\rel$
can in some cases constrain the location of the lens.  \citet{ob161928}
combined the high value of $\mu_\rel$ for OGLE-2016-BLG-1928 with the
{\it Gaia} source proper motion to argue that the lens was almost
certainly in the disk. Given the low value $\theta_\e=0.84\,\muas$,
this implied a very low lens mass $M = 0.23\,M_\oplus(\pi_\rel/125\,\muas)^{-1}$.

Third, by measuring the proper motion, one obtains a definite estimate
of the wait time until AO observations can distinguish between FFP
and 2L1S interpretations.  For example, using the fiducial parameters
of Equation~(\ref{eqn:delay}), and keeping in mind that giant sources
require 1.5 FWHM separation, the six FSPL FFP candidates found to date have
first observation epochs of
OGLE-2012-BLG-1323 (2027),
OGLE-2016-BLG-1540 (2024),
OGLE-2016-BLG-1928 (2024),
KMT-2017-BLG-2820 (2028),
OGLE-2019-BLG-0551 (2039), and
KMT-2019-BLG-2073 (2032).


The launch of the {\it Nancy Grace Roman} (f.k.a.\ {\it WFIRST}) satellite
will provide another path to FFPs.  \citet{johnson20} estimate that
FFP population models consistent with the \citet{mroz17} short-$t_\e$ events
would lead to several hundred {\it Roman} detections (see their Figure 7).
For the events among these that are generated by wide-separation planets
(as opposed to genuine FFPs), a substantial fraction of the hosts will
be directly detected as blended flux.  This is because most of the
sources are M dwarfs, and hence have comparable flux to the lens hosts,
while the fields are relatively sparse at {\it Roman} ($\sim 100\,\mas$)
resolution.  However, for those events without measurable blended flux $f_b$
(either because $f_b$ is small or the errors in $f_b$ are large
due to the faintness of the source and the small number of magnified
points), then ground-based 30m AO will still be required to confirm
that these are FFPs.  Because the sources are small, most will not have
$\mu_\rel$ estimates.  Thus, adopting a relatively conservative
$\mu_\rel > 1.5\,\masyr$ limit, and using the scaling of
Equation~(\ref{eqn:delay}), one should wait $\sim 15\,$yr after the
mission, i.e., circa 2045 before vetting these candidate FFPs.
Nevertheless, a fraction $\sim \rho=\theta_*/\theta_\e\sim
(0.25\,\muas)/(5\,\muas)= 5\%$ will have $\mu_\rel$ measurements,
meaning that of order a dozen FFP candidates can be vetted by 2035.
Thus, space-based and ground-based FFP surveys will remain complementary
for several decades.

Finally, we note that the frequency of wide orbit planets can be studied
by looking for short ``bumps'' in the long-term light curves of
archival microlensing events \citep{ob110173,poleski20}.  One could
then check whether this population of wide-orbit planets was large enough
to account for the rate of FFP candidates.  If we assume that the
source must come within $u_0< u_{0,\rm lim}$
planetary Einstein radii to be detected,
and that a fraction $\xi$ of a given microlensing light curve is covered,
then the probability of detection is
\begin{equation}
p = {4 u_{0,\rm lim}\over 2\pi} \xi {\sqrt{q}\over s} =
1.0 \times 10^{-4} {u_{0,\rm lim}\over 0.7}
{\xi \over 0.2}
\biggl({ q\over 3\times 10^{-5}}\biggr)^{1/2}
\biggl({ s\over 5}\biggr)^{-1}.
\label{eqn:probwide}
\end{equation}
Here, we have scaled to the planet-host mass ratio $q$ that would
be appropriate if the FFP candidates turn out to be bound and
to the light-curve coverage factor that would be appropriate for
such short events that are observed from a single site.
If we consider only the $N_{\rm ev}\sim 5000$ OGLE-IV events
in fields with the necessary $\Gamma \geq 1\,{\rm hr}^{-1}$ cadence, and
assume $N_{\rm pl}=5$ planets per star, then the expected number of detections
$N_{\rm ev}N_{\rm pl}p=2.5/(s/5)$, which could provide marginal evidence
for the wide-orbit hypothesis (if detected).  Note that for
Jupiter-mass planets ($q\sim 2\times 10^{-3}$, $\xi\sim 0.7$)
the expectation is much higher: $N_{\rm ev}N_{\rm pl}p=14\,N_{\rm pl}/(s/5)$.
Moreover, events in lower-cadence fields could also be probed.
To date, four wide ($s>3$) bound planets have been found by microlensing,
with $(s,\log q)=[(4.7,-1.5),(4.4,-1.8),(4.6,-3.3),(5.3,-3.6)]$
\citep{ob160263,mb12006,ob110173,ob08092}.  As noted by \citet{ob110173},
there is a 1.5 dex gap in $q$ between the first two and last two,
although one should keep in mind that the sample is not homogeneously
selected.

Here we report on a new FSPL FFP candidate, which
was discovered by applying the above-described supplemental giant-source search
to the 2017 KMT light-curve data base.  This search returned 232 microlensing
candidates, of which 15 had not previously been identified in the
EventFinder search, which had yielded 2817 candidates.  Being the third on this
list of 15, we designate it KMT-2017-BLG-2820, following the convention
introduced by \citet{ob161928} for OGLE-2016-BLG-1928.  We also discuss
the broader implications of the accumulating set of FSPL FFP discoveries.

{\section{{Observations}
\label{sec:obs}}

KMT-2017-BLG-2820 occurred at (R.A., Decl)$_{J2000}=$
(17:34:58.25, $-28$:32:51.22), corresponding to $(l,b)=(-0.91,+2.18)$.
It therefore lies in KMT field BLG14, which was observed at the
time of the event at nominal cadences of
$\Gamma = (1.0,0.75,0.75)\,{\rm hr}^{-1}$ from KMT's three observatories
at the Cerro Tololo Interamerican Observatory (KMTC), South African
Astronomical Observatory (KMTS), and Siding Springs Observatory (KMTA),
respectively.  Each facility has a 1.6m telescope equipped with a
$2^\circ\times 2^\circ$ camera.  Most observations were in Cousins $I$.
In 2017, every tenth $I$-band observation from KMTC was complemented
by an observation in Johnson $V$ band, while this applied to only
every twentieth observation from KMTS and KMTA.

The event also lies in OGLE field BLG653, which was observed in Cousins $I$ band
with a cadence of $\Gamma=0.17\,{\rm hr}^{-1}$ from OGLE's 1.3m telescope
at Las Campanas Observatory, which is equipped with a $1.4\,{\rm deg}^2$
camera.  OGLE also took occasional $V$-band images.  Unfortunately,
neither KMT nor OGLE took such images when the source was sufficiently
magnified to measure its color.

Neither KMT nor OGLE alerted the event in real time, so there
was no possibility of follow-up observations.  We checked and found
that the UKIRT microlensing survey \citep{ukirt17} was taking observations
close to the peak of the event.  Unfortunately, however, while this field
was in their 2016 footprint, it was not in their 2017 footprint.
Because UKIRT observes in $H$ and $K$, even a single such observation
would have yielded a very good color measurement.

Data reductions were carried out using specific implementations of
difference image analysis \citep{tomaney96,alard98}, by \citet{albrow09}
for KMT and \citet{wozniak2000} for OGLE.

{\section{{FSPL Analysis}
\label{sec:fspl}}

Figure~\ref{fig:lc} shows the color-coded data from the four observatories
together with the best-fit zero-blending FSPL model, which has four
parameters (apart from the flux parameters).  These are the
three \citet{pac86} parameters $(t_0,u_0,t_\e)$ and $\rho$,
where $t_0$ is the time of closest approach and $u_0$ is the
impact parameter in units of $\theta_\e$.  The fit parameters
are given in Table~\ref{tab:1L1S}.  This will be our preferred solution.
However, in contrast to the cases of some other FSPL FFP candidates,
there is no compelling reason from the light curve data themselves to conclude
that the source is unblended.  For example, for OGLE-2019-BLG-0551,
the blending was poorly constrained, but the source color was well measured
to be similar to that of the baseline object.  This implied
that strong blending was unlikely.  But, more importantly, it implied
that the $\theta_\e$ determination was independent of the blending
\citep{ob190551}.  The current case is closer to that of KMT-2019-BLG-2073,
for which the source color was not measured and the blending
fraction $\epsilon\equiv f_b/f_{\rm base}$ was measured to only about
$\sigma(\epsilon)\sim 20\%$ at the $1\,\sigma$ level \citep{kb192073}.
However, in the present case, while there is also no color measurement
and the $1\,\sigma$ limit on $\epsilon$ is similar, there is a
strong $3\,\sigma$ limit $\epsilon<0.4$ that implies that the
source definitely dominates the light from the baseline object.
This fact will play an important role in the argument given in
Section~\ref{sec:blend} that the source is most likely not blended.

Before continuing, we remark on the technical point that we
implement ``zero blending'' by fixing $f_{s,\rm KMTA} = f_{\rm base, KMTA}$
i.e., we equate the source and baseline fluxes at KMTA.  We find
in Section~\ref{sec:cmd} that the color and magnitude offsets of the source
from the clump are nearly identical for KMTA and OGLE (and indeed are
similar for all four observatories).
So, from this standpoint, either (or really, any) observatory
could be used.  However, the $\rho$ measurement depends primarily on
the KMTA data, and $\theta_\e$ is directly proportional to the
square root of the normalized surface brightness
$\hat S\equiv f_s/\rho^2$ \citep{ob190551,kb192073}.  Therefore,
it is really only the fixing of $f_{s,\rm KMTA}$ that directly impacts the result.
We considered fixing some or all of the other source fluxes, but this
does not significantly change the values of the other parameters
compared to just fixing $f_{s,\rm KMTA}$.

Table~\ref{tab:1L1S} also shows the parameters for the case of free blending.
The estimate of the blended flux is consistent with zero, and
the remaining parameters have similar values to the zero-blending fit.
However, the errors are much larger.  Nevertheless, the normalized
surface brightness $\hat S$ has a fractional error
of only 3.6\%, implying that this measurement contributes only
1.8\% to the uncertainty in $\theta_\e=\sqrt{\hat S}\times$[color term].
That is, as discussed in some detail by \citet{kb192073}, as regards
the crucial measurement of $\theta_\e$, the real uncertainty introduced
by unknown blending is the degree to which it implies that the source
color differs from that of the baseline object, which is used
in the analysis as a proxy for the source color.  To address this
issue further requires the analysis of two types of auxiliary data:
photometric and astrometric.

\section{{CMD}
\label{sec:cmd}}

Figures \ref{fig:cmd_ogle} and \ref{fig:cmd_kmta} show color-magnitude
diagrams (CMDs) within a $2^\prime\times 2^\prime$ box centered on
the event for OGLE and KMTA, respectively.
In each case, the ``baseline object'' is shown in black,
while the clump centroid is shown in red.  The OGLE CMD is calibrated,
and the KMTA CMD has been shifted by offsets derived from relatively
bright comparison stars, $14<I_{\rm OGLE}<16.9$.  We need to compare these
two CMDs because, while the OGLE photometry is unquestionably better
(see, e.g., the lower giant branches in the respective figures),
the normalized surface brightness $\hat S \equiv f_s/\rho^2$ is best
constrained from the KMTA data.

We measure the offset from
the clump $\Delta[(V-I),I] = [(V-I),I]_{\rm base} - [(V-I),I]_{\rm clump}$, finding
$(+0.07,-0.19)$ and $(+0.07,-0.18)$ for OGLE and KMTA, respectively.
That is, even though the OGLE photometry is substantially better,
the KMTA photometry is adequate for measuring this offset.

In the zero-blending model, the ``baseline object'' is the source.
Then, using the known dereddened position of the clump
$[(V-I),I]_{\rm clump,0}=(1.06,14.50)$ \citep{bensby13,nataf13}, we obtain
$[(V-I),I]_{s,0}=(1.13,14.31)\pm(0.03,0.05)$,
where the principal source of error is from
centroiding the clump.  We cannot use a substantially larger area
because of differential reddening.  Then employing the standard
procedure of \citet{ob03262}, we convert to $[(V-K),K]$ using the
color-color relations of \citet{bb88} and apply the color/surface-brightness
relation of \citet{kervella04} to obtain,
\begin{equation}
\theta_* = 7.05 \pm 0.44\,\muas,
\label{eqn:thetastar}
\end{equation}
where we have added 5\% to the error, in quadrature, to account for
systematic errors in the overall method. Using the zero-blending parameters
from Table~\ref{tab:1L1S}, we then obtain
\begin{equation}
\theta_\e = {\theta_*\over\rho} = 5.94 \pm 0.37 \muas,
\qquad
\mu_\rel = {\theta_*\over t_*} = 7.95 \pm 0.52\, \masyr.
\label{eqn:thetae+mu}
\end{equation}

We re-emphasize that Equations~(\ref{eqn:thetastar}) and
(\ref{eqn:thetae+mu}) only apply under the assumption
of zero blending.  However, from the standpoint of
measuring $\theta_\e$, it is equally important to
emphasize that this measurement is affected by blending
only to the extent that the blend color differs from
that of the baseline object.  As discussed in Section~\ref{sec:fspl},
$\theta_\e = \sqrt{\hat S} \times$[color term].  Because $\hat S$ is
nearly invariant, $\theta_\e$ is unaffected by blending, provided
that it does not change the estimated color of the source.

\section{{Astrometry}
\label{sec:astrom}}

If there is blended light that is displaced from the source by
$\Delta\btheta_b$, then the source position (measured from difference
images near peak) will be displaced from the baseline object by
\begin{equation}
\Delta\btheta_s \equiv \btheta_s-\btheta_{\rm base} = -\epsilon\Delta\btheta_b,
\label{eqn:dthetas}
\end{equation}
where $\epsilon = f_b/f_{\rm base}$ is the fraction of the baseline-object
flux that is due to the blend.  Using the three good seeing images near
peak, we measure, in $0.4^{\prime\prime}$ pixels,
and in the (West,North) coordinate system of the detector,
$\btheta_s=(156.005,145.177)\pm(0.014,0.016)$, where the error
bars are the standard errors of the mean of the three measurements.
The baseline object position is $\btheta_{\rm base}=(155.990,145.210)$.
While we do not have an independent way to estimate the error bars
of this latter measurement, we judge them to be of the same order
as those of $\btheta_s$ because the baseline object is bright and
isolated and because the baseline flux is similar to the difference
flux at peak.  Together, these yield
\begin{equation}
\Delta\btheta_s(N,E) = (-13.2,-6.0) \pm (9.1,7.9)\,\mas.
\label{eqn:dthetas2}
\end{equation}
Even assuming Gaussian statistics (which would be somewhat optimistic),
this has a probability of $p=26\%$ under the hypothesis that the true
value is zero (i.e., either $f_b=0$ or $\Delta\theta_b=0$).  Hence,
the astrometric measurement does not provide positive
evidence in favor of blended light, and it is consistent with
zero blended light.  We now turn to the limits
and constraints on blended light.

{\section{{Three Types of Blend}
\label{sec:blend}}

As discussed in Section~\ref{sec:fspl}, we have adopted the parameters
of the zero-blend fit, even though the light curve permits 20\% blending
at $1\,\sigma$ and 40\% at $3\,\sigma$.  In this section, we justify
this choice.

Logically, there are only three possible sources of blended light:
a companion of the source, a companion of the lens, and an ambient
star that is unrelated to the event. We consider these in turn.

{\subsection{{Companion of the Source}
\label{sec:compsource}}

First we note that the light curve provides only weak constraints on
a putative source companion.
The Einstein radius is smaller than the source (i.e., $\rho>1$),
so a putative companion would not be magnified during the event and
would have an extremely low probability of being magnified before
or after the event.  If the source companion were sufficiently close, it
could give rise to a xallarap signal, of which there is no evidence.
Because the source is a giant, with $R_*\sim 12\,R_\odot$, a companion could give
rise to ellipsoidal variations over the entire light curve, provided
that the source companion were at separations less than
a few tenths of an astronomical unit.
These are the only constraints on this scenario from the light curve.

Second, the astrometric measurement likewise provides only weak constraints.
If a source companion contributed significantly to $f_{\rm base}$,
which is the only case of interest here, and if it were widely separated
from the source, then it would induce an offset between the source
and baseline object, which is not seen.  For example, for a separation
of $1300\,\au$ (i.e., a period of $P\sim 10^7\,$days) and $\epsilon=0.3$,
this would lead to an offset $\Delta\theta_s\sim 50\,\mas$, in
contradiction to Equation~(\ref{eqn:dthetas}).  This implies that
the combination of photometric and astrometric constraints leaves
open the vast majority of the binary-separation distribution for
solar-mass stars \citep{dm91}.

However, the prior probability of such a companion is very
low, although not completely negligible.  To contribute at least $\epsilon>10\%$
of the baseline light, the companion would have to be on the lower
giant branch (or possibly in the clump).  The baseline-object CMD
position corresponds to a roughly solar mass star, and these spend
less than 1 Gyr on the lower giant branch, compared to about 10 Gyr
on the main sequence.  According to Table~7 of \citet{dm91}, less
than 10\% of solar type stars have companions with mass ratios of
0.75 -- 1.00.  Therefore, less than 1\% of bulge giant stars on the upper
giant branch will have companions on the lower giant branch.

Although this probability is very low, we nevertheless now examine
the consequences of such a companion for the measurements of $\theta_\e$
and, secondarily, $\mu_\rel$.  The main point is that the lower
giant branch (and clump) have very similar colors to the baseline
object.  We therefore begin by asking how these parameters would
be affected if the colors were identical.  As already noted,
because $\hat S$ is an invariant, the value of $\theta_\e$ is
basically unaffected.  Then, because $t_*$ is also an invariant,
$\mu_\rel = \theta_*/t_*$ scales directly as $\theta_*$, i.e.,
$\mu_\rel = (f_s/f_{\rm base})^{1/2}\mu_{\rel,0}=\sqrt{1-\epsilon}\mu_{\rel,0}$,
where $\mu_{\rel,0}$ is the value derived in Section~\ref{sec:cmd}
for the zero-blending case.  For example, for $\epsilon=0.3$,
the proper motions would be slower by a factor
$\sqrt{1-\epsilon}\rightarrow 0.84$.  The major concern raised by
such an overestimated $\mu_\rel$  would
be that, via Equation~(\ref{eqn:delay}), one should really wait a factor 1.2
times longer before doing AO observations to search for a wide host.
Because there will not be any additional information that would rule
out such a source companion prior to AO observations,
this would mean that one should just
wait the extra time (or simply discount the $<1\%$ probability that
there is such a companion).

However, in fact, for $\epsilon<0.3$, the source companion
would be at least 1.3 mag below the baseline object, and so
directly below the clump, which is on average about
$\eta =0.07\,$mag  bluer than the baseline object in $(V-I)$.
Although $\eta$ is a logarithmic quantity, it is small
enough that we can treat it as linear in order to get an understanding
of its role.  Then $\Delta(V-I)_b \simeq \eta/(1-\epsilon)$, and thus
the source is $\Delta(V-I)_s \simeq \eta*\epsilon/(1-\epsilon)$
redder than the baseline object.  And this implies that
\begin{equation}
\Delta\ln\theta_* = -0.5\Delta\ln S(V-I)_0
= -0.5{d\ln S\over d(V-I)_0}\Delta(V-I)_s =
-0.5{d\ln S\over d(V-I)_0}{\epsilon\over(1-\epsilon)}\eta,
\label{eqn:dlnthetastar}
\end{equation}
where $S(V-I)_0$ is the source surface brightness as a function of color.
We evaluate $d\ln S/d(V-I)_0 = -1.83$ using the same method that
was used in Section~\ref{sec:cmd}, and thus obtain
\begin{equation}
\Delta\ln\theta_* \rightarrow 0.064\,{\epsilon\over (1-\epsilon)}
{\eta\over 0.07}.
\label{eqn:dlnthetaastar2}
\end{equation}
The first point is that the effect is small: for $\eta\sim 0.07$
and $\epsilon\la 0.3$, $\Delta\ln\theta_* \la 3\%$, which is less than the
statistical error.  Second, the impact on the estimated proper motion
is opposite in sign from the one identified above when we approximated
the source and baseline-object colors as being the same.  The combined
effect is approximately given by
\begin{equation}
\Delta \ln\mu_\rel \rightarrow {0.5\epsilon\over 1-\epsilon}
\biggl({d\ln S\over d(V-I)_0}\eta -1\biggr)
\rightarrow
 {\epsilon\over 1-\epsilon}\biggl(0.064{\eta\over 0.07} -0.5\biggr).
\label{eqn:dlnmurel}
\end{equation}
Thus, the color term only slightly mitigates the color-free term,
i.e., by of order $0.064/ 0.5=12\%$.

In summary, there is a very small $(<1\%)$ probability that the
source has a companion with sufficient flux to impact the determinations
of $\theta_\e$ and $\mu_\rel$.  If it does, it changes $\theta_\e$ by
substantially less than the statistical error.  The fractional
change in $\mu_\rel$ is larger, but still less than 15\%.  This
might lead one to increase the wait time for AO followup observations,
if one were sufficiently concerned about this $<1\%$ probability.

{\subsection{{Companion (i.e., Host) of the Lens}
\label{sec:complens}}

We will conduct a search for a host of the planet and thereby place
constraints on such a host in Section~\ref{sec:2l1s}.  However, from
the present perspective, all that is important about this search is
that there will be substantial parameter space, in particular,
in the domain of planet-host separation, that is unconstrained.

As we will show immediately below, it is a priori unlikely that
the lens contributes to the light of the baseline object at even
the $\epsilon=0.1$ level.  Nevertheless, the major concern is that the
very presence of such a host would prevent its detection in AO followup
observations by inducing an underestimate of the wait time.  In that case,
a non-detection would falsely lead to the conclusion that the lens
was an FFP\footnote{Note that a BD could also yield a non-detection, provided
that $\pi_\rel <\theta_\e^2/\kappa (13\,M_J) = 0.34\,\muas$, corresponding
to $D_S - D_L < 24\,$pc.  While not impossible, this is very unlikely.}.
However, we will show that there is, in fact, no basis for this concern.

The first point is that if the host is contributing $\epsilon>0.1$
of the baseline-object light, then it must be relatively nearby.
Comparing the observed position
of the clump (Figure~\ref{fig:cmd_ogle}) to its intrinsic position
\citep{bensby13,nataf13}, we derive $[(V-I),A_I]=(2.47,2.95)$.
At $D_L=(1,2,3,4)\,\kpc$, we estimate that the lens would lie
in front of $(35,50,70,85)\%$ of the dust. Then, to generate $\epsilon>0.1$,
the lens absolute magnitude would be $M_I<(8.8,6.7,5.3,4.3)$.  This
excludes essentially all lenses in the bulge and at $D_L\ga 4\,\kpc$ in
the disk from contributing significantly to the blended light, as well
as excluding the great majority at somewhat smaller distances.

To understand why blending from the remaining possible lenses cannot
undermine the wait-time estimate, we first consider the special case
that the (observed) lens color is the same as that of the baseline object.
The proper motion will then be overestimated by a factor $(1-\epsilon)^{-1/2}$
so that true separation will be (1.42,1.25) FWMH for $\epsilon=(0.1,0.3)$,
rather than 1.5 FWHM.  But the flux ratio in $I$ (and so in $K$,
because the colors are the same), will be $\epsilon/(1-\epsilon)=(0.11,0.43)$.
The first would easily be resolved at 1.2 FWHM, while the second would
easily be resolved at 1.0 FWHM.  See, for example, Figure~1 of
\citet{ob05071c}.  For lenses that are bluer than the source, the
$K$-band flux ratio
will be somewhat reduced compared to this estimate.  For example,
for a solar-like star at $D_L=(3,4)\,\kpc$, $\eta\sim(1.2,0.8)$,
leading to flux ratios that are a factor roughly (3,2) smaller in $K$ band
relative to $I$ band.  However, in these cases, the proper motion
will not actually be underestimated because the source is
substantially redder than the baseline object.  On the other hand,
if the source were redder than the baseline object
because the lens was an extremely nearby late M dwarf, e.g., $\eta\sim 0.5$,
then the proper motion could be underestimated by a factor (0.88,0.65)
for $\epsilon=(0.1,0.3)$, leading to true offsets of (1.3,1.0) FWHM.
However, these values would still be adequate even at the
$I$-band flux ratio $\epsilon$, and the $K$-band ratio
would be significantly higher.

The resilience of the wait-time estimate is due to the fact that
it was derived to enable lens-flux measurements in the face of
extreme flux ratios $\la 10^{-3}$, whereas
blending does not play a significant role unless $\epsilon\ga 10^{-1}$.

Therefore, there is no real possibility of failure of future AO
observations due to adopting the ``naive'' $\mu_\rel$ estimate given
in Section~\ref{sec:cmd}.


{\subsection{{Ambient Star}
\label{sec:ambient}}

The astrometric measurement in Section~\ref{sec:astrom} places strong
constraints on blends by ambient stars. We adopt a conservative upper limit,
$\Delta\theta_s <25\,\mas$, which leads to an upper limit on the offset
of an ambient star at $\Delta\theta_b = \Delta\theta_s/\epsilon$,
which covers an area $\Omega=\pi(\Delta\theta_s/\epsilon)^2$.
For $\epsilon>(0.1,0.2,0.3)$, the surface densities of stars with
$I<I_{\rm base}-2.5\log \epsilon = (19.76,19.01,18.57)$ are
$n=(0.073,0.045,0.035)\,{\rm arcsec}^{-2}$, with corresponding
probabilities $p=n\Omega=(143,22,8)\times 10^{-4}$.  Thus, the
probability of an $\epsilon>0.1$ ambient star is small (1.4\%),
while that of an $\epsilon>0.2$ ambient star is negligible.

{\subsection{{Summary}
\label{sec:summary}}

Among the three possibilities for blended light (companion of the source,
companion of the lens, and ambient star), two have both low probability
of existing and low impact if they do exist.  For both source companion
and ambient star, the probability is of order 1\% or less.  For source
companions, the only real scientific impact is that allowing for
this possibility suggests to extend the wait time for AO observations
by 20\%.  However, such observations can be taken at first AO light
on 30m telescopes, regardless.  For ambient stars, only $\epsilon\la 0.1$
blends have relevant $(\ga 1\%)$ probabilities,
and these have only a small impact on the
observables $\theta_*$ and $\mu_\rel$.

If there is
a lens companion, then most likely it has $\epsilon<0.1$.  For example
essentially all bulge lenses would have hosts with $\epsilon<0.02$,
essentially all disk lenses at $D_L>4\,\kpc$ would have hosts with
$\epsilon<0.1$, and a large fraction of more nearby hosts would
also have $\epsilon<0.1$.  In sum, even if there is a host, the
probability that it has $\epsilon>0.1$ is small.  Nevertheless,
it is quite easy to conjure scenarios of hosts that are above this
threshold, as we demonstrated in Section~\ref{sec:complens}.  However,
we also showed there that such hosts would be detected by late-time
AO observations regardless of
their impact on the proper-motion estimate that was evaluated in
Section~\ref{sec:cmd}.

{\section{{Source Proper Motion}
\label{sec:sourcepm}}

It is possible in principle to distinguish between bulge and disk
lenses by combining the scalar lens-source relative proper motion
$\mu_\rel$ (derived from the microlensing analysis) with the
vector source proper motion $\bmu_s$ derived from external sources.
For example, \citet{ob161928} found that the source star for
OGLE-2016-BLG-1928 had a proper motion almost identical
to the centroid of the neighboring bulge field stars.  Because
$\mu_\rel\sim 10\,\masyr$ in that case, the lens had to be moving
at $10\,\masyr$ relative to the mean bulge motion.  However, the
{\it Gaia} proper motion diagram showed that there are few, if any,
bulge stars with such proper motions.

We pursue a similar investigation here.  We use 10 years of OGLE-IV data,
which we align to {\it Gaia} using a technique that is described
by \citet{oiv-pm}.  We find that the proper motion of the baseline object is
\begin{equation}
\bmu_{\rm base}(E,N) = (-7.55,-2.71)\pm (0.42,0.60)\masyr,
\label{eqn:mubase2}
\end{equation}
where we have doubled the formal errors derived from the scatter
of the fit, based on tests performed by \citet{oiv-pm} in regions
where overlapping OGLE fields provide two independent measurements.
This measurement agrees with {\it Gaia} within the errors, but is
more precise.  Note that {\it Gaia} errors in bulge fields are
also underestimated by a factor of about two.  In view of the
small probability of significant blended light found in
Section~\ref{sec:blend}, we identify the source with the baseline object,
$\bmu_s = \bmu_{\rm base}$.

Figure~\ref{fig:pm} shows this measurement (red), together with
the proper motions of bulge field stars (black).  The figure is
rotated to Galactic coordinates.  In addition, for the first time,
we show all proper motions in the geocentric frame at the
time of the peak of the event, when Earth was moving
${\bf v}_{\oplus,\perp}(E,N) = (+28.90,-0.79)\,\kms$ relative to the Sun.
Thus, before
rotating the OGLE-IV heliocentric proper motions to Galactic coordinates
we first subtract
$\bmu_{\oplus,R_0} = {\bf v}_{\oplus,\perp}/R_0 = (+0.74,-0.02)\,\masyr$,
where $R_0=8.2\,\kpc$.

In this way, we ensure that the magenta circle that is centered
on the geocentric source proper motion, and represents the
$7.95\pm 0.73\,\masyr$ {\it geocentric} lens-source relative proper motion,
accurately predicts the range of allowed geocentric lens proper motions.
Note that to obtain the $1\,\sigma$ range of the predicted
$\bmu_l = \bmu_s + \bmu_\rel$ we have added in quadrature the errors
for $|\bmu_s|$ and $\mu_\rel$.
As can be seen, this range is quite consistent with the lens lying
in the bulge (black points).

We now ask whether this annulus is also consistent with the lens lying
in the disk.  The blue circles represent the mean geocentric lens
proper motion for disk lenses lying at 2 kpc (right) and 5 kpc (left).
The error bars reflect the velocity dispersions of stars at each
distance.  The blue curve connecting the blue circles shows the mean
proper motions at $2<D_L/\kpc<5$.
We have assumed dispersions
of $\zeta^{1/2}\times (28,18)\,\kms$ where $\zeta=\exp(D_L/2.5)\,\kpc$
is the ratio of the local surface density to the one in the solar
neighborhood.  We also assume an asymmetric drift of
$v_{\rm rot} - \sqrt{v_{\rm rot}^2 - \zeta(47\,\kms)^2}$, where
$v_{\rm rot} = 235\,\kms$ is the local rotation speed.  We take account
of the motion of the Sun relative to the local standard of rest (LSR),
$v_{\odot,\perp}(l,b) = (12,7)\,\kms$, as well as the instantaneous motion
of Earth.  The mean estimates for each distance are well displaced
toward lower $\mu(l)$ from the origin.
Three factors contribute to this.  First, the
Sun is moving at $+12\,\kms$ relative to the LSR in this direction.
Second, Earth's instantaneous motion is $+15\,\kms$ relative to the Sun
in this direction.  Third, the asymmetric drift of stars at these distances
is in the opposite direction.  In the latitude direction, Earth's strong
motion toward Galactic south overwhelms the small northerly
motion of the Sun.

Hence, the mean expected motion is displaced from the origin, though
which the magenta annulus directly passes.  Nevertheless, after taking
account of the velocity dispersions of the lens (error bars), the
lens is consistent with being in the disk at the $1\,\sigma$ level
and at any distance from us.
Thus, the proper-motion analysis is quite consistent with the lens
lying in either the bulge or the disk.

{\section{{2L1S Analysis}
\label{sec:2l1s}}

If the lens has a host, then it may leave its signature on the
(seemingly) 1L1S event, either by generating a second, much longer,
bump in the light curve or by creating caustic structures on the main,
short timescale, event.  To search for such host signatures,
we follow the procedures described by \citet{kb192073} for
KMT-2019-BLG-2073.  In particular, we add three parameters to the
fit $(s,q,\alpha)$, i.e., the planet-host separation in units
of the total-mass (i.e., host+planet) Einstein radius, the ratio
of the host and planet masses, and the angle of the host-planet axis
with respect to the lens-source relative motion.  We center the coordinate
system on the planetary caustic.  We conduct a grid search in these variables,
seeding the remaining four at their values implied by the 1L1S solution.
Figure~\ref{fig:grid} shows the result of this search, with a clear minimum at
about $(s,q)\sim (6,100)$, which is favored over the 1L1S solution by
$\Delta\chi^2 = -22$.  Figure~\ref{fig:lc_grid} shows the corresponding
light curve for the best fit.

To understand the origin of this $\chi^2$ improvement, we plot
the cumulative distribution of $\Delta\chi^2$ in Figure~\ref{fig:dchi2}.
This shows that the net $\chi^2$ improvement comes entirely from KMTA,
with the other three observatories canceling each other out.  Hence,
the most likely explanation for the improvement is low-level
systematics in the KMTA data.

Returning to Figure~\ref{fig:grid}, we see that all models
with $s<3$ and $q>100$ have $\Delta\chi^2>36$ relative to the minimum,
and so $\Delta\chi^2>14$ relative to 1L1S.  The same applies
to models with $s<2.5$ and $q>10$.  We regard such models as ruled out.
Thus, if future AO observations identify a host, yielding estimates
of $M_{\rm host}$ and $D_L$, and if second-epoch
observations measure the projected planet-host separation $a_\perp$ (see
Section~\ref{sec:wide}), then
$a_\perp > 3\,\theta_{\e,\rm host} D_L$.  For example, if $M_{\rm host}=0.8\,M_\odot$
and $D_L=6\,\kpc$, then $q\simeq 90$, and we predict
$a_\perp>9\,\au$.  As we discuss in Section~\ref{sec:wide}, this threshold
is near the limit with current instrumentation (and modest efforts),
but plausibly could be achieved with 30m AO.

{\section{{Discussion}
\label{sec:discuss}}

{\subsection{{Nature of the Observed FFP Population}
\label{sec:nature}}

KMT-2017-BLG-2820 is the sixth FSPL FFP candidate discovered to date.
Five of these six (all except OGLE-2016-BLG-1928, which has much smaller
$\theta_\e$ and $t_\e$) have Einstein
radii in the range $2.4 < \theta_\e/\muas < 9.2$.  While these
five were not selected homogeneously, they do have some common
features that should help us to understand their parent population.
First, all five events occurred on giant-star sources
with angular radii $5.4<\theta_*/\muas <20$.
Second, all five have Einstein timescales $3.7<t_\e/{\rm hr}<9.1$.

This Einstein-timescale range can be directly compared
to that of the six PSPL FFP candidates discovered by \citet{mroz17},
$3.1 < t_\e/{\rm hr} < 8.0$.  On this basis,
these two samples appear to be drawn from the same underlying population.
In the Appendix, we show that the two samples have consistent discovery
rates.

We can express
the definition of $\theta_\e$ (Equation~(\ref{eqn:tedef})) as a relation
scaled to a value of $\pi_\rel$ that is typical of bulge lenses:
\begin{equation}
M = 0.20\,M_J
\biggl({\theta_\e\over 5\,\muas}\biggr)^2
\biggl({\pi_\rel\over 16\,\muas}\biggr)^{-1}.
\label{eqn:massscale}
\end{equation}
Hence, if the five FSPL FFPs lay in the bulge, this population would
consist of sub-Jovian
gas giants and ice giants.  However, even though bulge lenses generally
dominate the microlensing event rate, one must be cautious about this
interpretation.  It is possible, for example, that nature produces
very few gas-giant FFPs (or wide separation planets), in which case
these low-$\theta_\e$ lenses would mostly or all be in the Galactic
disk, with correspondingly lower masses.  That is,
Equation~(\ref{eqn:massscale}) can equally be written as,
\begin{equation}
M = 8.2\,M_\oplus
\biggl({\theta_\e\over 5\,\muas}\biggr)^2
\biggl({\pi_\rel\over 125\,\muas}\biggr)^{-1}.
\label{eqn:massscale2}
\end{equation}

One might hope to distinguish between these alternatives based
on the measured (scalar) proper motions, $\mu_\rel$.
However, in all five cases, $\mu_\rel$
is consistent with either a bulge- or disk-lens interpretation.
In principle, if $\mu_\rel\ga 10\,\masyr$, then a measurement of $\bmu_s$
that put it near the center of the bulge proper-motion distribution
can effectively rule out a bulge lens \citep{ob161928}.  However, for
the only one of these five events with $\mu_\rel>10\,\masyr$,
OGLE-2016-BLG-1540, Figure~3 of \citet{ob161540} shows that the proper motion
of the lens  is in fact consistent with it
being in either the disk or the bulge.

{\subsection{{The Einstein Desert}
\label{sec:desert}}

Another way to potentially distinguish between the super-Earth/disk
and sub-Jovian/bulge hypotheses would be to analyze the full
$\theta_\e$ distribution from a homogeneously selected FSPL sample.
Under either hypothesis, the observed paucity of short-$t_\e$/small-$\theta_\e$
events\footnote{There is only one (OGLE-2016-BLG-1928) out of a total of 12.}
is explained by declining sensitivity, even if the underlying
population of FFPs were rising toward lower mass.
However, the two hypotheses make different predictions for the
long-$t_\e$/large-$\theta_\e$ tail of the FFP $\theta_\e$ distribution.
If these events are due primarily
to sub-Jovian/bulge FFPs, then (assuming similar planet formation and
evolution in the disk and bulge\footnote{Note that in order to contradict
the logic of this argument, there would have to be some mechanism that
enhanced the production of wide-orbit or unbound sub-Jovian planets in the
bulge relative to the disk.  For example, the suggestion of
\citet{thompson13}, that gas-giant formation may be suppressed in the bulge
due to the harsh radiation environment, would work in the opposite direction.
However, \citet{mctier20} have shown that $\sim 10\%$ of bulge stars suffer
a stellar encounter within $10\,\au$ during
their 10 Gyr lifetime.  Some, but far from
all of these would disrupt sub-Jovian gas giants in the cold outer regions.
In the impulse approximation, the planet would gain
$\delta v = 2 GM/bv = 1\,\kms(M/M_\odot)/[(b/10\,\au)(v/200\,\kms)]$,
which is far below its orbital velocity.  Here, $M$ is the mass of
the perturber, $v$ is its speed relative to the planet, and $b$ is the
impact parameter.  Thus, to eject the planet (to a wide orbit or out of
the system), the star would have
to come within $\sim 1\,\au$ of the planet, for which the probability
is 100 times smaller.}) there should also be a population
of sub-Jovian/disk
FFPs that give rise to events with roughly
$\sqrt{125\,\muas/16\muas}\sim 2.8$ times larger $t_\e$ and $\theta_\e$, i.e.,
centered on $t_\e\sim 0.7\,$days  and $\theta_\e\sim 14\,\muas$, respectively.
However, $\theta_\e\sim 14\,\muas$ is very nearly at the (logarithmic) center
of the ``gap'' found by \citet{kb192073} in the interval
$4.8<\theta_\e/\muas<31.7$, which we dub the ``Einstein desert''.
And  $t_\e\sim 0.7\,$days is close to the minimum
of Figure~1 of \citet{mroz17}.

In contrast to the sub-Jupiter/bulge hypothesis, the Einstein desert is a
natural consequence of the super-Earth/disk hypothesis, which
predicts that there are no ejected planets (whether arriving in
very wide or unbound orbits) that are more massive than super-Earths.
In this scenario, there are no events with intermediate $t_\e$ or $\theta_\e$
from the disk (due to non-existence of sub-Jovian FFPs),
and very few (or no) FFP events from the bulge (due to the $\theta_\e$ and
$t_\e$ of super Earths having very low detection sensitivity).

While the \citet{mroz17} PSPL sample was selected according to
homogeneous criteria, the five FSPL events that we have just
been discussing are inhomogeneously selected.  However, \citet{kb192073}
presented a plan for collecting a homogeneous sample of giant-source
FSPL events, within which the FFP subsample could be analyzed.
They applied this approach to 2019 KMT data and found two FFP candidates
among their 13 FSPL events.  KMT-2017-BLG-2820 was found by applying
the same approach to 2017 and 2018 data.  It is therefore the
third FFP candidate found in this developing homogeneous sample.  The analysis
of the 2017--2018 FFP events is ongoing, and it will be extended to 2016
as well.  If the Einstein desert remains parched in this expanded sample
of FSPL events (as indicated by a preliminary analysis),
then this will tend to confirm the super-Earth/disk
hypothesis.  On the other hand, if this ``desert'' were gradually populated by
intermediate $\theta_\e$ lenses, then the sub-Jovian/bulge hypothesis
would gain traction.

{\subsection{{Free-Floating versus Wide-Orbit Planets}
\label{sec:wide}}

As with all FSPL FFPs candidates, one can distinguish between the
FFP and wide-orbit scenarios for KMT-2017-BLG-2820
by imaging the system at high resolution
when the source and lens are sufficiently separated to be resolved.
From Equations~(\ref{eqn:delay}) and (\ref{eqn:thetae+mu}), this
will be in 2028 or 2026 for Keck AO observations in $K$ or $H$,
respectively.  Or one might be slightly more conservative to allow
for the measurement errors in $\mu_\rel$, or concerns about systematic
errors in $\mu_\rel$ due to unrecognized blending.  However, we have
argued that the latter concern is minor.

If the planet has a host, then the host should appear in these observations
with a separation approximately given by $\Delta\theta = \mu_\rel\Delta t$,
where $\Delta t$ is the elapsed time since the event.  It will then
be possible to determine or constrain the planet-host projected
separation by taking a second observation several years later \citep{gould16}.
That is, suppose that the two measurements of the $\Delta \btheta$
offset between the host and source are taken
$\Delta t_1$ and $\Delta t_2$ after the event, each with precision $\sigma$.
Then the host position at the time of the event is given by
\begin{equation}
\Delta\btheta_0 =
{\Delta\btheta_1\Delta t_2 - \Delta\btheta_2\Delta t_1
\over\Delta t_2 - \Delta t_1}
\pm {\sqrt{(\Delta t_1)^2 + (\Delta t_2)^2}\over \Delta t_2 - \Delta t_1}\sigma.
\label{eqn:dtheta0}
\end{equation}
For example, for $\sigma=500\,\muas$ (e.g., \citealt{mb13220b}),
$\Delta t_1= 10\, $yr, $\Delta t_2= 16\,$yr,
and a lens distance of $D_L=6\,\kpc$, the error corresponds to $10\,\au$.
This approach only works for planet-host separations
that are small compared to the FWHM, i.e., 55 mas for our fiducial parameters,
corresponding to $330\,\au$ for $D_L=6\,\kpc$.  For separations that are
of order the FWHM, the host will appear in some random position that
is inconsistent with the one predicted from $\Delta\theta_1 =\mu_\rel\Delta t_1$.
Then the host would be identified as such because it moved with the
proper motion derived from the microlensing fit between the two epochs.
At sufficiently large separations, the method becomes limited by
a background of stars moving with similar proper motions.  \citet{gould16}
discusses the problem of distinguishing among ``Kuiper, Oort, and
Free-Floating Planets'' in greater detail.

On the other hand, if no host is seen, this does not absolutely
prove that the lens is an FFP.  As mentioned, it could be a BD
that passed $D_S - D_L<24\,\pc$ in front of the source.  Or it could
have a dark host, such as a BD, old white dwarf, neutron star, or
black hole.  However, while these rare exotic systems might explain
one non-detection, they could not explain an ensemble of non-detections.

While there is no purely empirical evidence that would distinguish between
the FFP and wide-orbit explanations for the FFP candidates, the
balance of evidence from a combination of theoretical arguments
and observational data strongly favors the FFP hypothesis.

First, as we have described above, the existence of the ``Einstein desert''
implies that these lenses are super-Earths in the disk rather than
gas giants in the bulge.  While this desert must be confirmed,
we can report that with most of the FSPL analysis of 2017-2019 complete,
the desert remains.

Second, to account for their six detected PSPL FFP candidates,
\cite{mroz17} required 5--10 times more FFP super-Earths than stars.
Given the typical lower limits $s\ga 3$ on hosts, adopting
typical $D_L\sim 4\,\kpc$ distances for disk lenses, and adopting
$a_{\rm snow} = 2.7\,\au (M/M_\odot)$ for the snow line, this would
imply projected separations, $a_\perp\ga 2.5\,a_{\rm snow}(M/0.3\,M_\odot)^{-1}$,
relative to the snow line, i.e., beyond the orbit of Jupiter in the
Solar System.  There could easily be one super-Earth in this
zone, but if there are 5--10 per star, then they must be spread to
considerably larger orbits.  For the Solar System, it is well established
that the timescales in these regions are too slow to form super-Earths.

Therefore, in the wide-orbit hypothesis,
the super Earths must have formed closer in and then been
``ejected'' from these inner regions to where they are seen today.
There is, indeed, the well worked out ``Nice model''
\citep{nice} for such
an ejection to explain the present positions of Uranus and Neptune.
This model must (and does) explain how these planets retained
their roughly circular orbits, but it relies on a Jupiter-Saturn
resonance.  Such two-gas-giant systems are relatively rare
\citep{gould10,wittenmyer20}.  Hence,
more generally, ``mass expulsions'' (i.e., 5--10 planets)
would take place by planet-planet scattering, or pumping (as created the
Oort cloud) that would send the planets into highly elliptical
orbits.  If the mechanism were scattering, it would require fine tuning
to have the planets lose most of their energy, but remain bound.  This already
implies that FFP and Oort-like orbits should predominate over
Kuiper-like orbits.  Moreover, in the process of the super-Earth migration
to pumping orbits, it seems likely that most would be scattered
out of the system.

As we have emphasized, it will be possible to test these conjectures
by AO followup to locate hosts, and by second-epoch AO to determine
their separations.






{\section{{Conclusion}
\label{sec:conclude}}

We have discovered a new FSPL FFP candidate, KMT-2017-BLG-2820,
with Einstein
radius $\theta=5.94\pm 0.37\,\muas$ and lens-source relative proper
motion $\mu_\rel = 7.95\pm 0.52\,\masyr$.  Whether this is truly
a ``free-floating planet'' or is simply a very wide-separation planet can
can be determined with excellent (though not perfect) confidence
by AO followup observations made before the end of the current decade.
If the latter, then the planet-host projected separation can be
measured with roughly $10\,\au$ precision from a second AO epoch.

KMT-2017-BLG-2820 was discovered in an ongoing systematic search for
giant-source FSPL events within 2016--2019 KMT data
\citep{kb192073}.  It is the
third FFP candidate in this developing homogeneous sample and the sixth
FSPL FFP candidate overall.  Five of these six FSPL FFP candidates
have a very similar Einstein timescale distribution as the six PSPL FFP
candidates found by \citet{mroz17} in their study of 1L1S events
found in the OGLE-IV database.  Moreover, the detection rates of the
\citet{mroz17} PSPL sample and the sample being collected under the
\citet{kb192073} protocols are comparable (see Appendix).
We therefore argue that the
five FSPL FFP candidates and six PSPL FFP candidates are drawn from
the same population.  Based on the measured Einstein radii
$\theta_\e \sim 5\,\muas$ of the former, these could be either
sub-Jovian FFPs in the bulge or super-Earth FFPs in the disk.
We argue that, if the Einstein desert in the $\theta_\e$ distribution of
giant-source FSPL events tentatively found by \citet{kb192073}
is confirmed, then it argues for the latter scenario, which
was already suggested by \citep{mroz17} on different grounds.

In making our parameter estimates, we have adopted the zero-blending
model.  First, while the blending parameter $\epsilon=f_b/f_{\rm base}$
is relatively weakly constrained by the light curve at the $1\,\sigma$
level, $\epsilon = 0.12\pm 0.10$, it has a firm upper limit
$\epsilon<0.4$ at the $3\,\sigma$ level.  Thus, the source dominates
the baseline object, which sits on the upper giant-branch track.
Hence, the probability for $\epsilon>0.1$ from a source companion
is less than 1\% due to rarity of lower giant-branch stars.  The
probability for $\epsilon>0.1$ from an ambient star is similarly
restricted by the close astrometric alignment between the source
and the baseline object.  If the lens has a host, then this
would supply blended light at some level.  However, essentially
all potential hosts at $D_L>4\,\kpc$ generate $\epsilon\ll 0.1$
and this applies to most potential hosts at $D_L<4\,\kpc$ as well. In any
case, we showed that ignoring such possible blended light from
the host would not cause one to overestimate $\mu_\rel$ and thereby
underestimate the wait time for AO followup observations.

\acknowledgments
We thank Subo Dong for stimulating discussions.
This research has made use of the KMTNet system operated by the Korea
Astronomy and Space Science Institute (KASI) and the data were obtained at
three host sites of CTIO in Chile, SAAO in South Africa, and SSO in
Australia.
Work by C.H.  was supported by the grants of NationalResearch
Foundation of Korea (2017R1A4A1015178 and 2019R1A2C2085965).
The OGLE project has received funding from the National Science Centre,
Poland, grant MAESTRO 2014/14/A/ST9/00121 to AU.
%
%

\appendix
\section{Relative Rates of the PSPL and FSPL FFP Samples}
\label{sec:append}

We show here that the detection rates underlying the six \citet{mroz17} PSPL
sample and the three FSPL events obtained so far under the program
outlined by \citet{kb192073} are roughly comparable.  Because \citet{kb192073}
have not yet characterized their selection function, we adopt
a basically empirical approach.  And because the Poisson errors of this
comparison are roughly $\sqrt{1/6 + 1/3}\sim 70\%$, there is not much
to be gained by detailed, highly precise calculations.  We therefore seek
only to demonstrate rough consistency.

\citet{mroz17} searched nine OGLE fields over about 5.5 seasons,
three with cadence $\Gamma \sim 3\,{\rm hr}^{-1}$ and
six with cadence $\Gamma \sim 1\,{\rm hr}^{-1}$.  They showed
that their detection efficiency in the timescale range of the actual
detections was about two times higher in the former.  Hence, if the
fields had equal underlying rates, there should be a nearly
equal number of detections
in the two sets of fields.  Instead, there are five and one.
Some of this difference is due to lower underlying rates in the
lower-cadence fields.  And some may be due to Poisson fluctuations.
Nevertheless, in order to maintain a homogeneous empirical approach,
we restrict attention to the three higher-cadence fields, with total
area $\Omega_{\rm OGLE} = 4.2\,{\rm deg}^2$.  For simplicity we label the
six events shown in Figure~3 and Table~1 of \citet{mroz17} as M-1 ... M-6.
We note that apart from M-1 (which has a giant source), all six have
$u_0<0.6$.  We therefore estimate the effective cross section as
$2\times 0.7\times \langle\theta_\e\rangle \rightarrow 8.4\,\muas$,
where we have used
$\langle\theta_\e\rangle = 6\,\muas$ based on the ensemble of
FSPL FFPs (excluding OGLE-2016-BLG-1928, which has a timescale outside
the range of either the \citealt{mroz17} sample or that of
the \citealt{kb192073} approach).  We estimate the source
absolute magnitudes using $I_s$ from Table~1, extinctions from
\citet{gonzalez12} (with $A_I = 7\,A_K$),
and Galactic-bar distances from \citet{nataf13}.
Apart from M-6 (which is also the only event with $(1-\epsilon)\ll 1$),
the events all have $M_I < 3.3$.  We judge this to be the range
of source sensitivity.

Based on Figure~8 of \citet{kb192073}, we estimate that the FSPL FFP
sample has sensitivity to sources $M_I<0.5$.  From their Figure~4,
the cross section for the FSPL events is $2\,\theta_*$.  From Figure~8
(and keeping in mind that for a clump giant, $\theta_*\simeq 6\,\muas$)
we adopt $\langle\theta_*\rangle = 7\,\muas$ and so a cross section
of $14\,\muas$.  As discussed by \citet{kb192073}, and illustrated
by their Figure~5, their FSPL search is about equally sensitive
in all fields $\Gamma \ga 1\,{\rm hr}^{-1}$.  This is fundamentally
because giant-source events have a full duration of about 20 hours.
However, as also shown by their Figure~5, detections are dominated
by events near the Galactic plane.  Although extinction is higher in the
northern bulge, this does not affect the relative detection rate much, again
because the sources are giants.  We judge the effective area of the
search to be $\Omega_{\rm KMT}=10\,{\rm deg}^2$.

To estimate the relative detection rates, we combine four factors:
1: effective number of sources, 2: effective area, 3: effective cross
section, 4: mean diurnal time coverage.  We estimate ratios
(FSPL/PSPL) $=(1/10)(10/4.2)(14/8.4)(2.5)=1.0$.  This can be compared
to the actual rates of 3/(3 yr) and 5/(5.5 yr) for FSPL and PSPL
respectively.  The estimate of the first factor is based on
the \citet{holtzman98} luminosity function (and the effective $M_I$ limits).
The second and third factors were described above.  The fourth factor
derives from the fact that OGLE is operating from a single site, while
KMT is operating from three sites, one of which (KMTS) has somewhat
worse weather and one of which (KMTA) has substantially worse weather.

The fact that the ``predicted'' FSPL/PSPL ratio (1.0) and the observed
ratio (1.1) are nearly identical should not be over-interpreted: the comparison
has no physical meaning below the Poisson errors.  In addition, our
estimates have considerable uncertainties (although below the Poisson
errors).  The only aim of this Appendix has been to demonstrate that
the two samples are consistent in terms of detection rate.

\begin{deluxetable}{lcc}
\tablecolumns{3} \tablewidth{0pc} \tablecaption{\textsc{1L1S
models}} \tablehead{ \colhead{Parameters} & \colhead{1L1S} &
\colhead{1L1S [$f_S(\textrm{KMTA})$=2.06]}  } \startdata
  $\chi^2/\rm{dof}$             &1994.017/1994         &1994.645/1995        \\
  $t_0$ $(\rm{HJD}^{\prime})$   &7910.046 $\pm$ 0.009  &7910.043 $\pm$ 0.007 \\
  $u_0$                         &0.164 $\pm$ 0.105     &0.302 $\pm$ 0.053    \\
  $t_{\rm E}$ $(\rm{days})$     &0.288 $\pm$ 0.015     &0.273 $\pm$ 0.006    \\
  $\rho$                        &1.096 $\pm$ 0.079     &1.187 $\pm$ 0.009    \\
  $t_*$ $(\rm{days})$           &0.314 $\pm$ 0.010     &0.324 $\pm$ 0.006    \\
  $\hat{S}$                     &1.503 $\pm$ 0.054     &1.465 $\pm$ 0.023    \\
  $f_S({\rm KMTA})$             &1.805 $\pm$ 0.204     &2.06                 \\
  $f_B({\rm KMTA})$             &0.252 $\pm$ 0.204     &-0.003 $\pm$ 0.001   \\
  $f_S({\rm OGLE})$             &1.736 $\pm$ 0.165     &1.939 $\pm$ 0.085    \\
  $f_B({\rm OGLE})$             &0.244 $\pm$ 0.165     &0.042 $\pm$ 0.085    \\

\enddata
\tablecomments{$t_*\equiv\rho t_\e$ and $\hat{S}\equiv f_S/\rho^2$
are derived quantities and are not fitted independently.}
\label{tab:1L1S}
\end{deluxetable}

\begin{figure}
\plotone{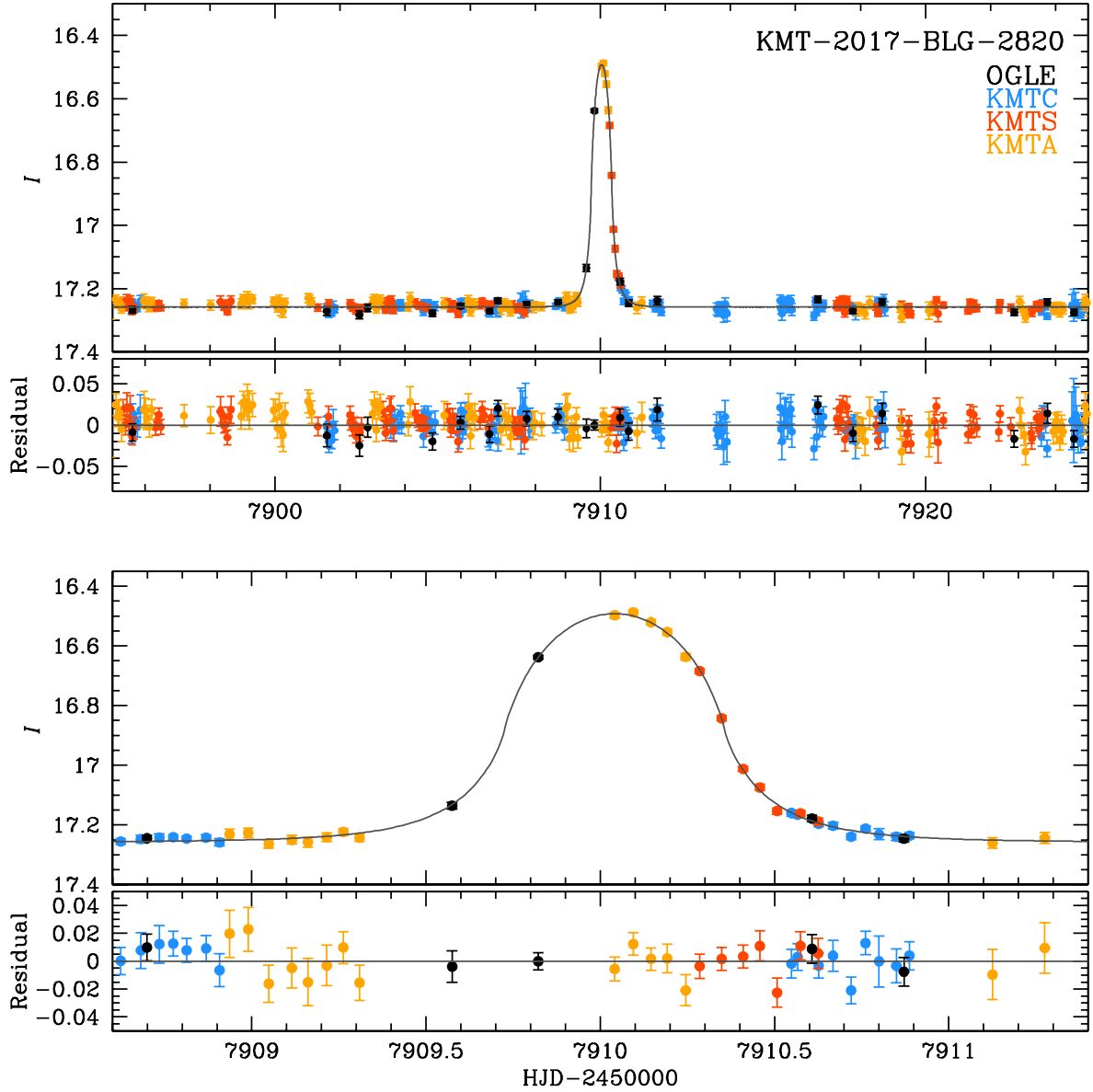}
\caption{Light curve and FSPL model for KMT-2017-BLG-2820, with
the source flux fixed to that of the ``baseline object''.
}
\label{fig:lc}
\end{figure}

\begin{figure}
\plotone{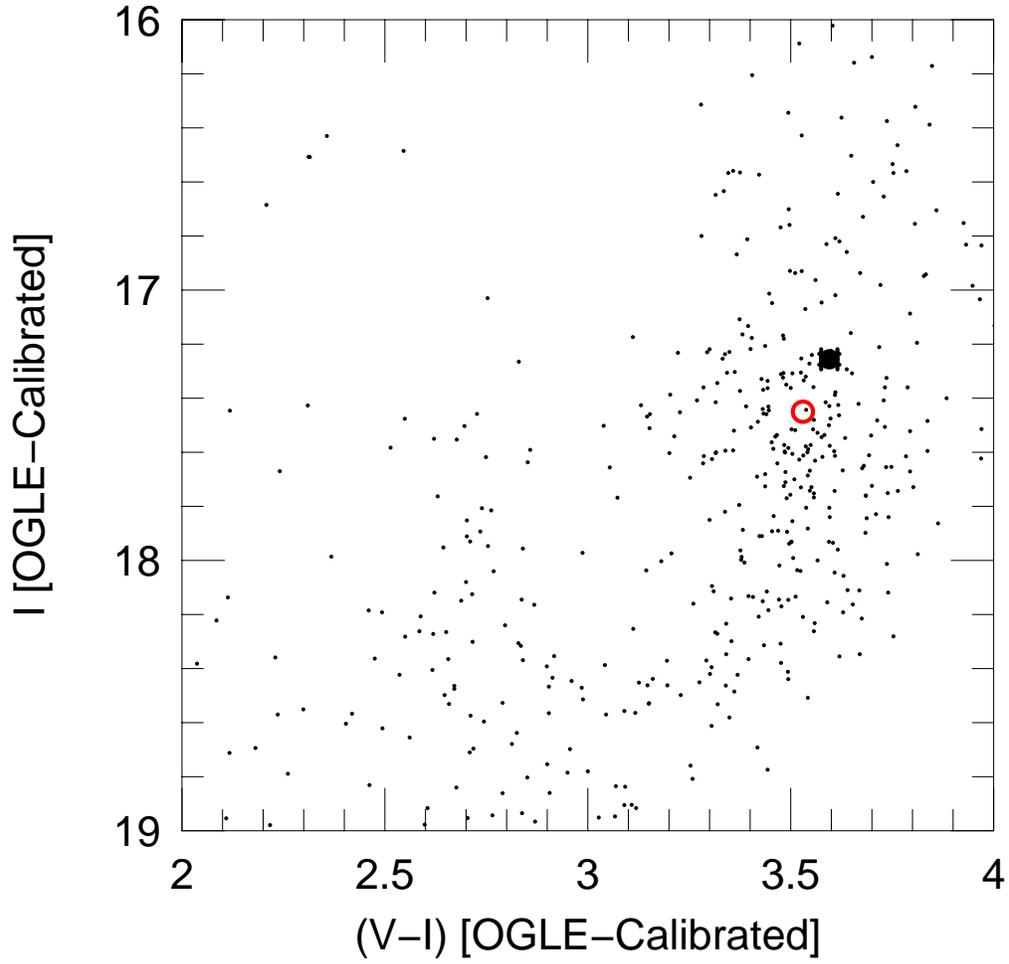}
\caption{Calibrated color-magnitude diagram (CMD) based on OGLE-IV data.
The black point is the baseline object, and the red circle is the
clump centroid.  The baseline object is either on the upper giant
branch or is a clump star that is superposed upon it.
}
\label{fig:cmd_ogle}
\end{figure}

\begin{figure}
\plotone{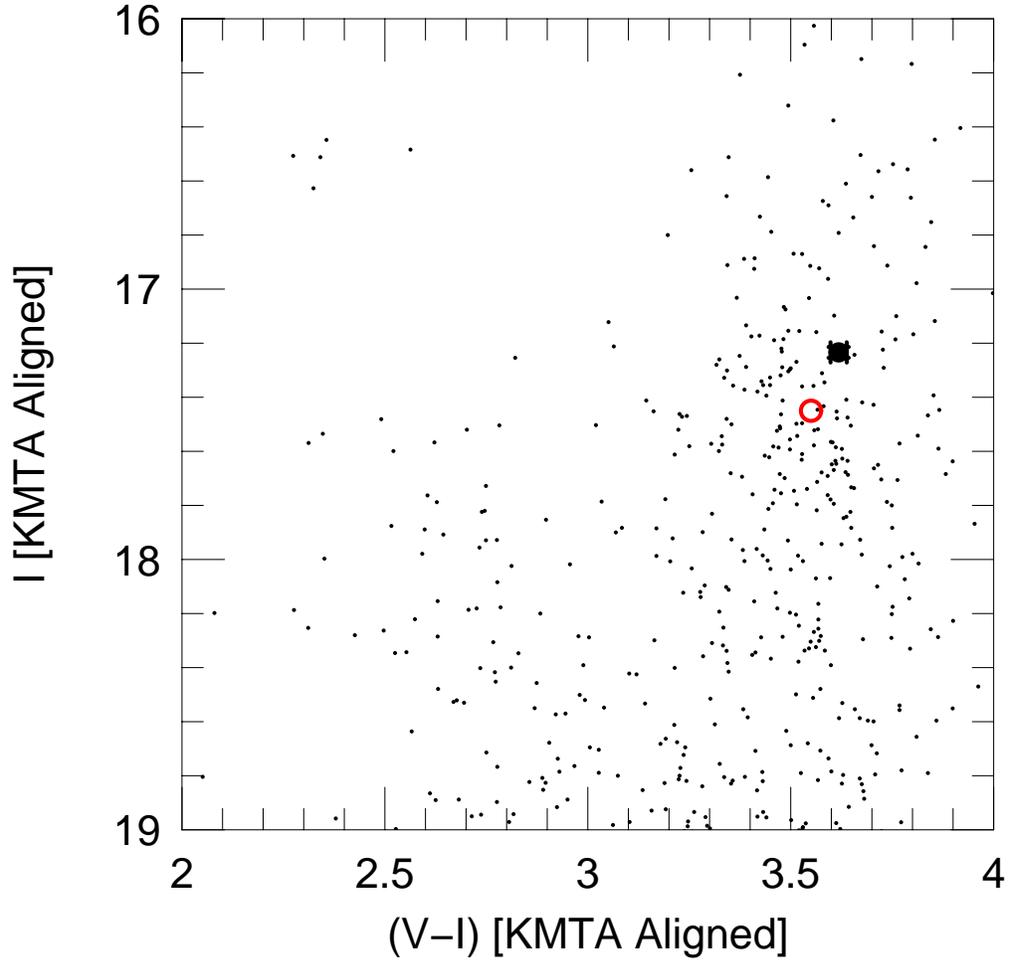}
\caption{CMD based on KMTA data.  Similar to Figure~\ref{fig:cmd_ogle}
except the axes have been aligned to those of Figure~\ref{fig:cmd_ogle},
based on the offsets of bright stars $14<I_{\rm OGLE}<16.9$.  The
resulting offset of the baseline object (black) and clump
centroid (red) are almost identical to Figure~\ref{fig:cmd_ogle},
}
\label{fig:cmd_kmta}
\end{figure}

\begin{figure}
\plotone{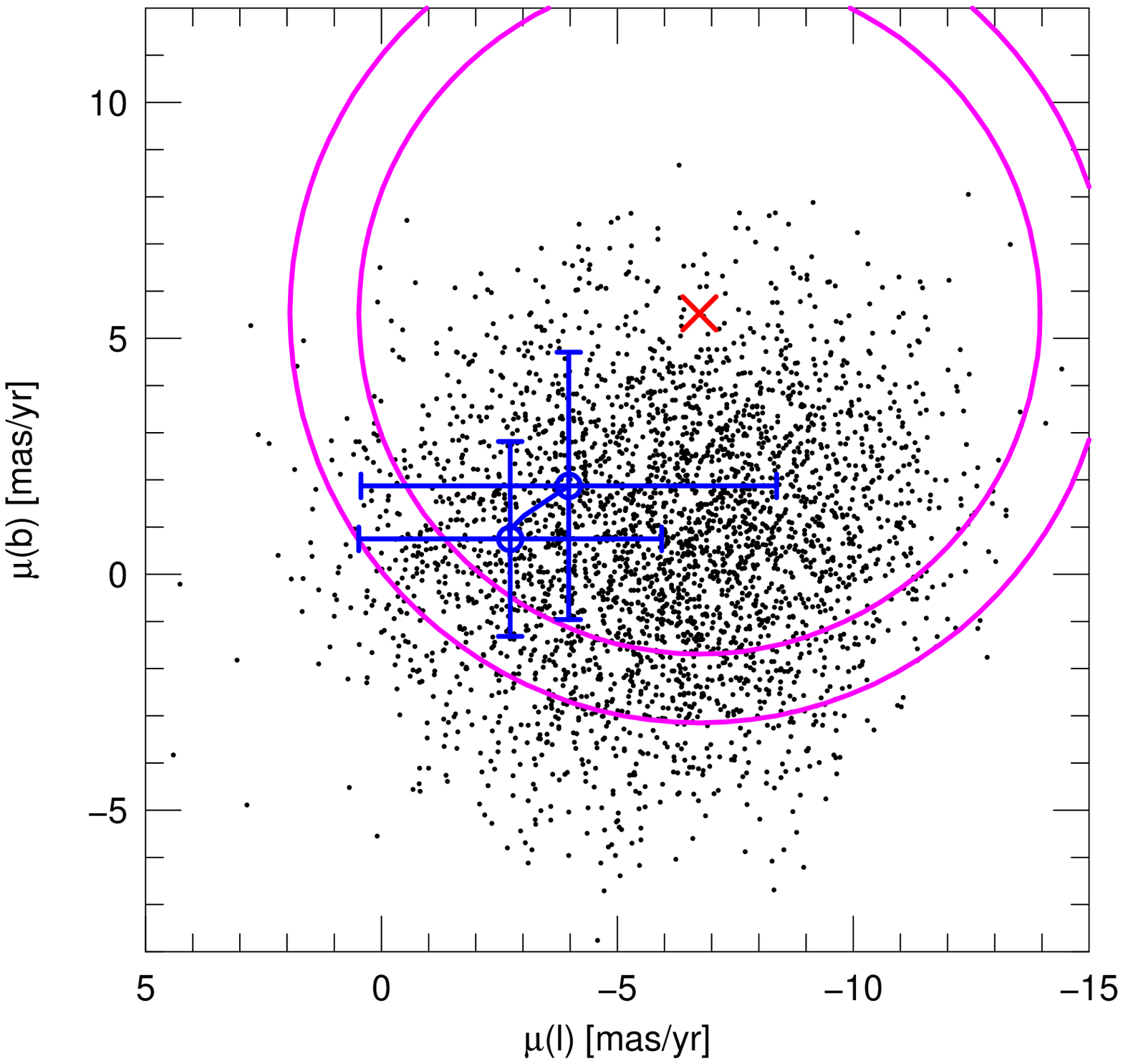}
\caption{OGLE-IV proper-motion diagram of the KMT-2017-BLG-2820 field,
with bulge red giants and clump giants shown in black and the microlensed
source shown in red.  The magenta annulus shows the $1\,\sigma$
range of the allowed lens proper motion $\bmu_l = \bmu_\rel + \bmu_s$,
given the measurements of $|\bmu_\rel|$ and $\bmu_s$ in
Equations~(\ref{eqn:thetae+mu}) and (\ref{eqn:mubase2}).
This allowed region can be compared to the predicted $\bmu_l$
for bulge (black dots) and disk (blue circles with error bars) lenses.
The latter are shown for 2 kpc (right) and 5 kpc (left), with a blue
curve showing the mean value at intermediate distances.  The
error bars represent the $1\,\sigma$ lens velocity dispersions
in each direction.  See text for details.  The lens is consistent
with either a disk or bulge location.
}
\label{fig:pm}
\end{figure}

\begin{figure}
\plotone{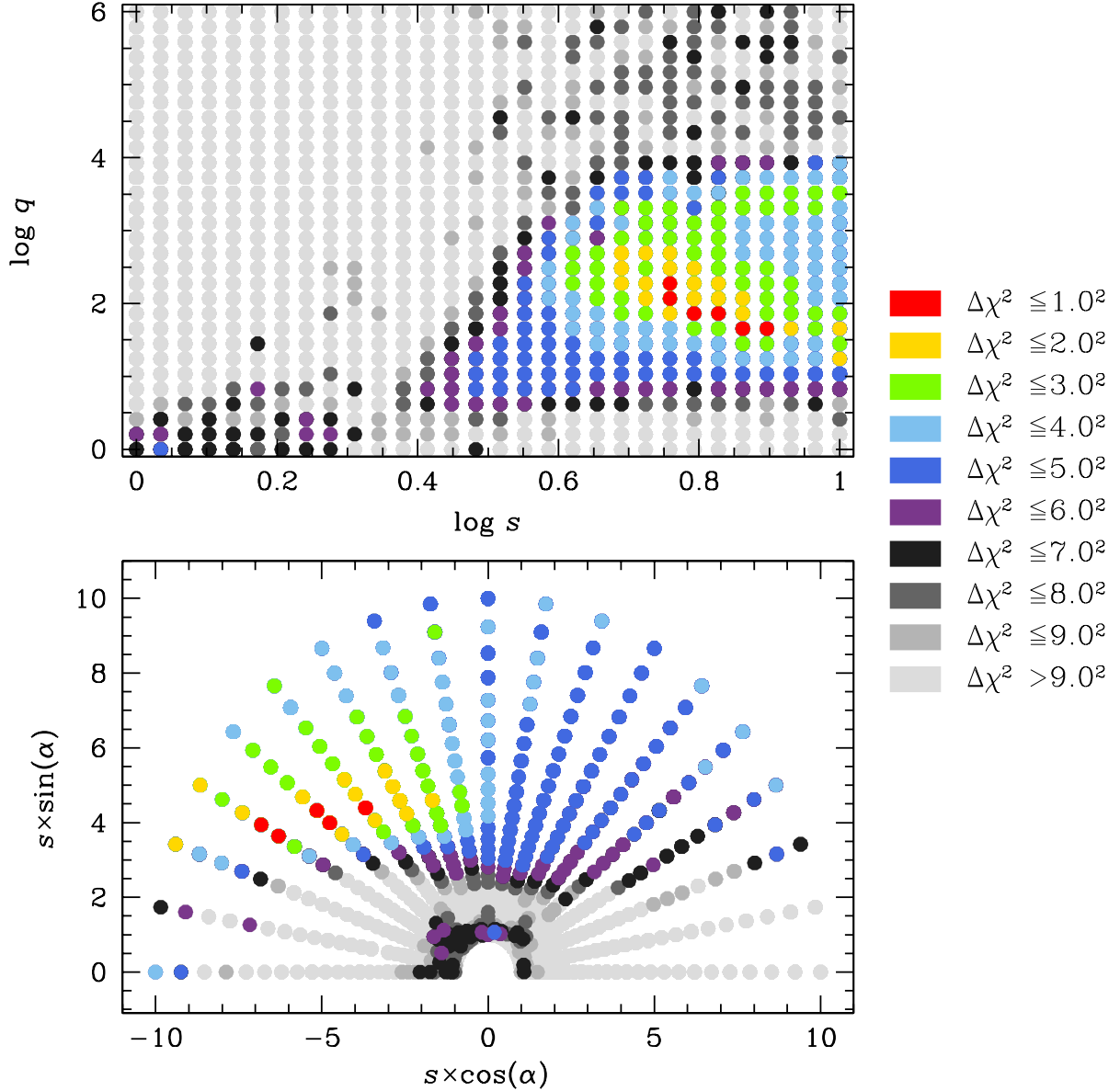}
\caption{Results of an $(s,q,\alpha)$ grid search for 2L1S models
of KMT-2017-BLG-2820.
There is a well-defined minimum at $(s,q)\sim (6,100)$ (upper panel),
with $\alpha$ at intermediate angles (lower panel), which has
a $\Delta\chi^2=22$ improvement
relative to the 1L1S model.  However, the improvement is due to low level
stellar systematics in the KMTA data,
not a real host of the FFP.  See Figure~\ref{fig:dchi2}.
We regard $\Delta\chi^2>36$
(so,$\Delta\chi^2>14$ relative to 1L1S) as ruled out.  This includes
$(s<3,q>100)\cup(s<2.5,q>10)$.
}
\label{fig:grid}
\end{figure}

\begin{figure}
\plotone{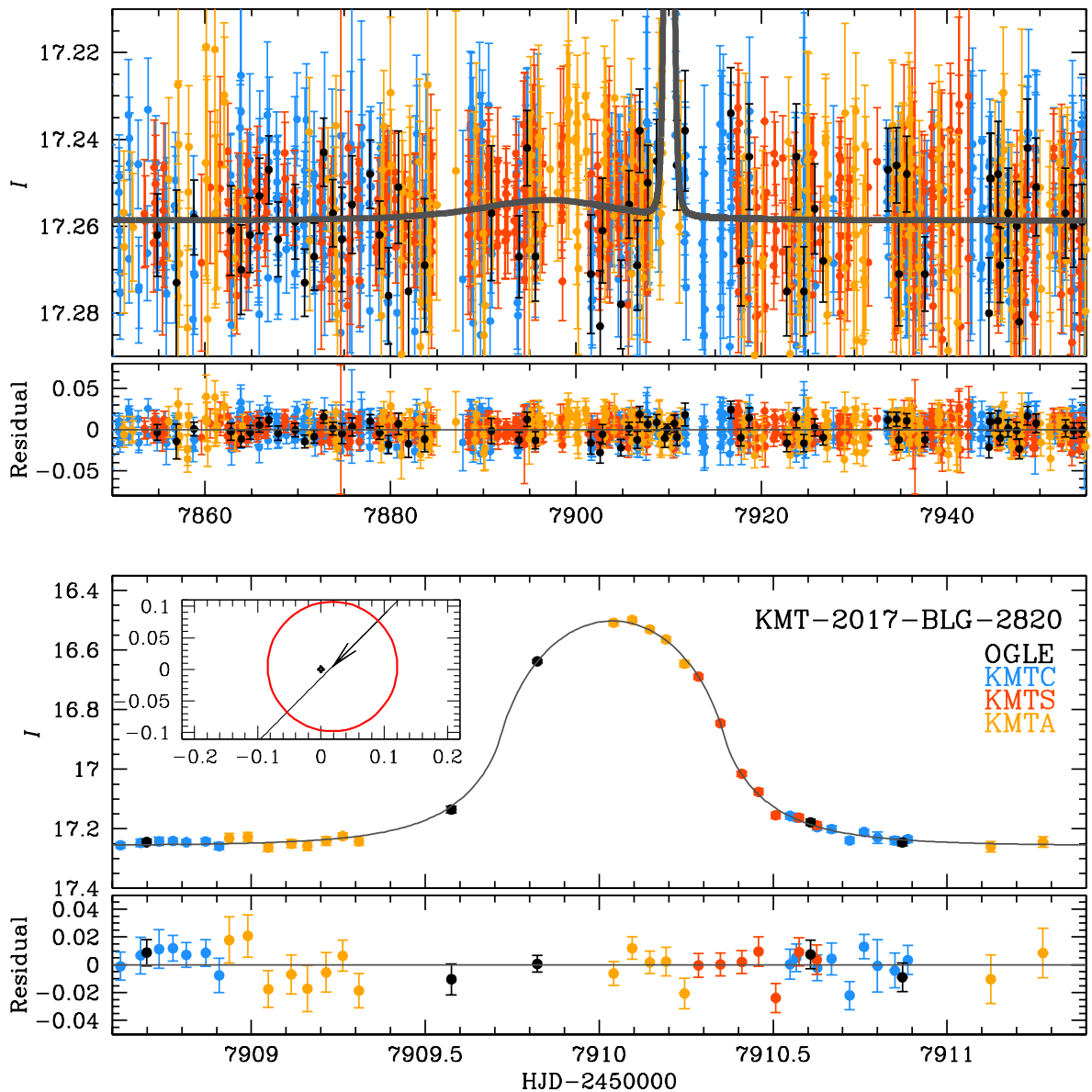}
\caption{KMT-2017-BLG-2820 light curve for the
best-fitting 2L1S model.  The putative host
would ``explain'' the $\sim 20$-day, 0.005 mag ``bump'' as the source passed
within $u_{0,host}\sim 4$ of the host.  However, the amplitude
of this bump is several times smaller than the error bars, and so
the bump requires further investigation.  See Figure~\ref{fig:dchi2}.
}
\label{fig:lc_grid}
\end{figure}

\begin{figure}
\plotone{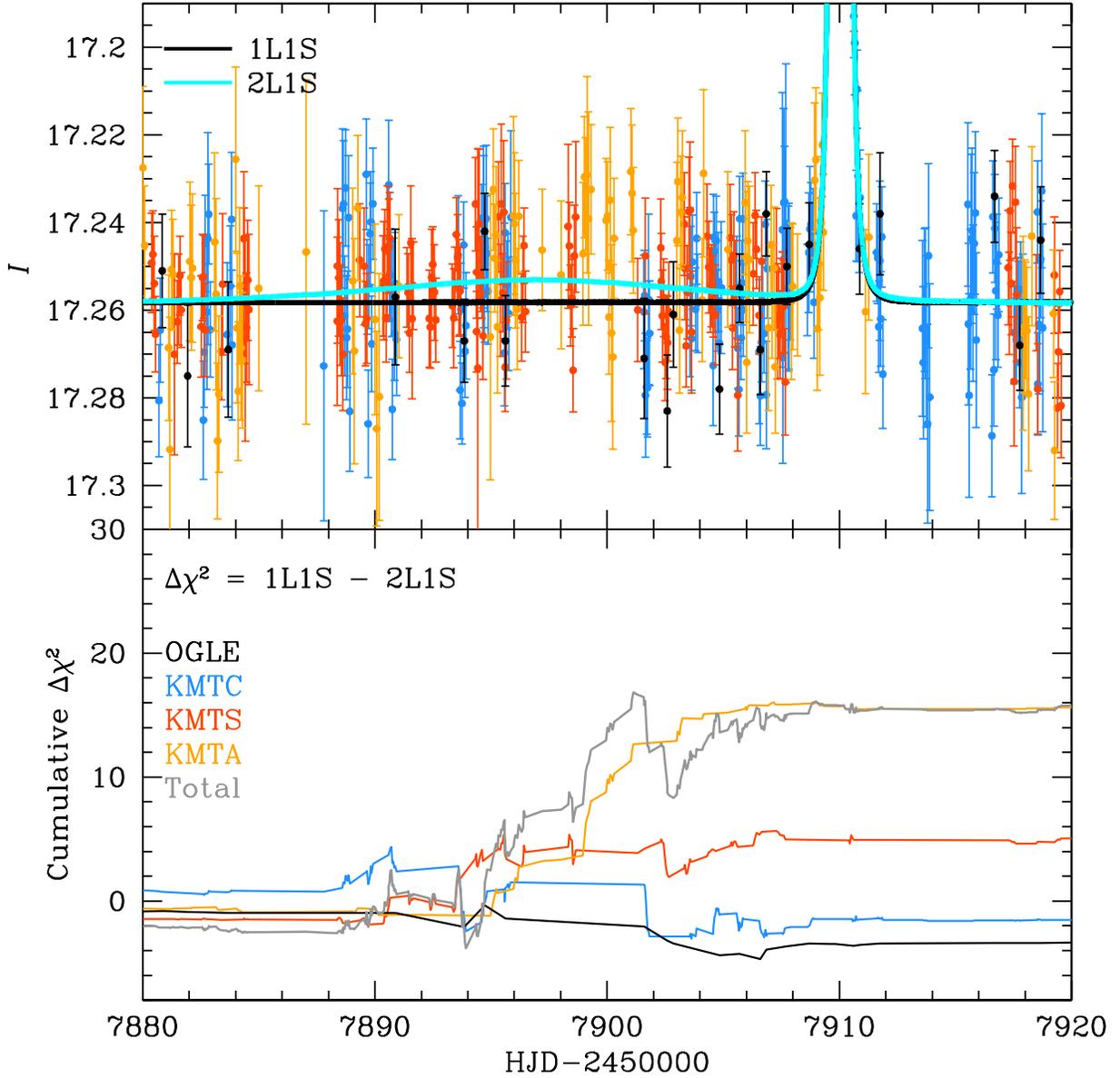}
\caption{Lower panel: Cumulative $\Delta\chi^2$ diagram, i.e., the sum
$\Delta\chi^2(t) = \sum_{t_i < t}[\chi^2_{\rm 1L1S}(t_i)-\chi^2_{\rm 2L1S}(t_i)]$
of the $\chi^2$ differences up to time $t$.  This shows that the
net signal comes entirely from KMTA data, with the contributions of the
remaining three observatories canceling each other out.
Upper panel: Data and 2L1S model, similar to Figure~\ref{fig:lc_grid}.
Comparing the two panels, we see that most of the ``signal'' comes from three
nights of KMTA observations, during which the data lie systematically above
the curve rather than generally matching the curve.  This is the
classic signature of a systematics-induced ``signal''.
}
\label{fig:dchi2}
\end{figure}


\end{document}